\documentclass[twocolumn,showpacs,preprintnumbers,amsmath,amssymb]{revtex4}

\usepackage{latexsym}

\usepackage[dvips]{color}
\usepackage{graphicx}
\usepackage{epsf}
\usepackage{dcolumn}
\usepackage{bm}

\usepackage{amssymb}
\usepackage{amsmath}
\usepackage{amsfonts}

\font\helv=cmssbx10

\def\beeta{\pmb{\eta}}

\begin{document}

\title{ Random cyclic matrices}

\author{Sudhir R. Jain}
\email{srjain@barc.gov.in}
\author{Shashi C. L. Srivastava}
\email{shashics@barc.gov.in}
\affiliation{Nuclear Physics Division, Van de Graaff Building,\\
 Bhabha Atomic Research Centre, Trombay, Mumbai 400 085, India}

\date{\today}


\begin{abstract}
We present a Gaussian ensemble of random cyclic matrices on the real field and study their spectral fluctuations.  These
cyclic matrices are shown to be pseudo-symmetric with respect to generalized
parity. We calculate the joint probability distribution
function of eigenvalues and the spacing distributions analytically and numerically. For small spacings, the level spacing
distribution exhibits either a Gaussian or a linear form. Furthermore, for the general case of two arbitrary
complex eigenvalues, leaving out the spacings among real eigenvalues, and,  among complex conjugate pairs,
we find that the  spacing distribution agrees completely with the Wigner distribution for Poisson process on a plane.
The cyclic matrices occur in a wide variety of physical situations, including disordered linear atomic chains and Ising model in
two dimensions. These exact results are also relevant to two-dimensional statistical mechanics and $\nu$-parametrized quantum chromodynamics.

\end{abstract}
\maketitle

With the pseudo-Hermitian extension of quantum mechanics \cite{bender,za,am}, it has become possible to
develop a number of new ideas, opening thereby interesting and important directions
of investigation. One of these advances has been in random matrix theory where pseudo-unitarily
invariant ensembles were presented \cite{aj} that exhibit completely different kind of level repulsion
as compared to the ensembles known  \cite{mehta,bipz,ginibre}. Thus, physical systems that violate parity and time-reversal
invariance (${\cal PT}$-symmetric)  exhibit level repulsion that could be linear or $\sim -S\log S$
where $S$ is the nearest-neighbour spacing of levels. However, an explicit analysis has been done only for
an ensemble of 2 $\times$ 2 matrices.

In this Letter, we present random matrix theory (RMT)  of $N \times N$ cyclic matrices with real elements. As we shall show, these matrices
are pseudo-symmetric with respect to ``generalized parity''.  Such matrices arise in very significant contexts,
the celebrated example being that of Onsager solution of two-dimensional Ising model \cite{onsager,kaufman}. They are encountered
in the treatment of linear atomic chains with Born-von K\'{a}rm\'{a}n boundary condition \cite{lowdin1} and
in understanding overlap matrices for molecules like benzene.  These matrices also occur as
transfer matrices in the theory of disordered chains \cite{borland} and in the general context of wave propagation in one-dimensional
structures \cite{luttinger}. In the latter example, generally, matrices of second order occur - thus, our earlier results \cite{jain} throw light
on the fluctuation properties of the eigenvalues.
Cyclic matrices also appear in the context of phase transitions in the spherical model \cite{spherical}. In all these
varied instances, as soon as there is a random parameter (e.g. external field or a random coupling in the example of Ising model), the level correlations
dictate the long time tails of
the time correlation functions which, in turn, relate to the relaxation of these systems when they are perturbed from
thermodynamic equilibrium \cite{srjpg}.

RMT appears in seemingly unrelated problems in physics and mathematics ranging from growth models, directed polymers, random sequences, to
Riemann hypothesis \cite{satya,ahmed,jgk}.
Also, the study of random matrices has been related to quantum chaos and exactly solvable models in a remarkable way \cite{guhr,jgk,srj}. Generically, the statistics
of spectral fluctuations of classically integrable, pseudointegrable, and chaotic systems follow respectively the
general features of Poisson, short-range Dyson model \cite{gj,ajk} or Semi-Poisson \cite{bgs}, and Wigner-Dyson ensembles \cite{bohigas}. However, for the
physical situations occurring in two-dimensional statistical mechanics where  time-reversal and parity  are violated \cite{halperin,frank,wen,kitazawa}, there is no general
understanding of the statistical nature of spectral fluctuations \cite{dj,jd}. Perhaps the first example of a billiard system  with a ${\cal PT}$-symmetric
(violating ${\cal P}$ and ${\cal T}$) Hamiltonian was a particle enclosed in a
rectangular cavity in the presence of an Aharonov-Bohm flux line \cite{djm}. For this classically pseudointegrable system, the spectral statistics of quantum energy levels was
found to exhibit level repulsion that is distictly different from the standard RMT \cite{mehta}. For these class of systems, an important step  was taken in \cite{aj}, and
the present work takes us to show the nature of these fluctuations in $N \times N$ cyclic matrices. The general case of $N \times N$ random
pseudo-Hermitian matrices remains open, however.

Let us consider an $N \times N$ cyclic matrix with real elements, $\{a_i\}$:
\begin{eqnarray}\label{eq:1}
\mbox{\helv M} = \left[\begin{array}{cccc}a_1&a_2&...&a_N\\a_N&a_1&...&a_{N-1}\\ \vdots & & & \\a_2&a_3&...&a_1\end{array}\right].
\end{eqnarray}
It is important to note that this matrix is, in fact, pseudo-Hermitian (pseudo-orthogonal) with respect to $\beeta $
\begin{eqnarray}\label{eq:2}
\beeta = \left[\begin{array}{cccccc}1&0&0&...&0&0\\0&0&0&...&0&1\\0&0&0&...&1&0 \\\vdots & & & & & \\0&1&0&...&0&0\end{array}\right],
\end{eqnarray}
that is,
\begin{equation}\label{eq:3}
\mbox{\helv M}^{\dagger} = \mbox{\helv M}^T = \beeta \mbox{\helv M} \beeta ^{-1}.
\end{equation}
Since $\beeta ^2 = $ identity, {\helv I}, $\beeta $ is introduced here as ``generalized parity". Thus, we have an
ensemble of random cyclic matrices (RCM)
that is pseudo-orthogonally invariant in the sense of (\ref{eq:3}). There are two distinct scenario with respect to time-reversal, ${\cal T}$ and parity, ${\cal P}$: (a) standard case where ${\cal T}$ and ${\cal P}$ are preserved, this case is trivially ${\cal PT}$-symmetric, and, (b) the case of ${\cal PT}$-symmetry where ${\cal T}$ and ${\cal P}$ both are broken. In case (a), one may study the fluctuations properties of energy levels after classifying the eigenfunctions according to definite parity (odd or even); however the case (b) belongs to a different class altogether. Whereas case (a) corresponds to the invariant ensembles of random matrix theory \cite{mehta}, case (b) has not been fully studied, only some partial results exist \cite{aj,jain} and RCM belong to this case. To our knowledge, the discrete
symmetries for operators represented by cyclic matrices are clearly spelt out here for the first time. Due to this generality, our final results are expected to be relevant for a wide variety of physical situations occurring in anyon physics \cite{halperin}, $\nu  $-parametrized quantum chromodynamics, fractional quantum hall systems \cite{jkjain}, etc.

The eigenvalues of {\helv M} are given by \cite{kowaleski}
\begin{equation}\label{eq:4}
E_{l} = \sum_{p=1}^{N} a_p \exp \frac{2\pi i}{N}(p-1)(l-1);
\end{equation}
($l = 1, 2, ..., N$), the maximum real eigenvalue being $ \sum_{i} a_i$. The diagonalising matrix is given by \cite{spherical}
\begin{equation}\label{eq:5}
U_{jl} = \frac{1}{\sqrt{N}} \exp \frac{2\pi i}{N}(j-1)(l-1).
\end{equation}
We consider a Gaussian ensemble of cyclic matrices with a distribution,
\begin{equation}\label{eq:6}
P(\mbox{\helv M}) \sim \exp - A~{\mathrm tr~} (\mbox{\helv M}^{\dagger} \mbox{\helv M})
\end{equation}
where $A$ sets the scale (of energy, for instance).

For the sake of simplicity, we present the analysis for an ensemble of $3 \times 3$ matrices.
We would like to obtain the joint probability distribution function (JPDF) of eigenvalues because all the correlations are related to it.
Also, we would like to show results on the spacing distribution as they enjoy a central place in discussions in quantum chaos, universality arguments,
and rule the dominant long-time tail in correlation functions.
We immediately see that ${\mathrm tr~} \mbox{\helv M}^{\dagger} \mbox{\helv M} = 3 (a_1^2 + a_2^2 + a_3^2)$. In effect, we
have $P(\{a_i\}) = \left(\frac{3A}{\pi }\right)^{\frac{3}{2}}e^{-3A\sum_{i} a_i^2}$. There are three eigenvalues - one real,
$E_1 = \sum_i a_i$
and a complex conjugate pair, $(E_2, E_2^*)$. We may define spacing as $S_{23} := |E_2 - E_3| = \sqrt{3}(a_3 - a_2)$ as well as
$S_{12} := |E_1 - E_2| = |\frac{3}{2}(a_2 + a_3) + \frac{i\sqrt{3}}{2}(a_2 - a_3)|$. Obviously, $S_{12} = S_{13}$. The JPDF of eigenvalues $P(\{E_i\})$ can be
written as
\begin{eqnarray}\label{eq:7}
P(E_1, E_2, E_2^*) = \left(\frac{A}{\pi} \right)^{\frac{3}{2}}e^{-A (E_1^2 + 2 |E_2|^2)}.
\end{eqnarray}
With this JPDF, spacing distributions can be found \cite{note}. Spacing distribution for the complex conjugate pair, $P_{cc}(S_{23})$
is given by
\begin{eqnarray}\label{eq:8}
P_{cc}(S_{23}) &=& \int \prod_{i=1}^{3}da_i P(\{a_i\})\delta (S_{23} - \sqrt{3}|a_3 - a_2|) \nonumber \\
&=& \sqrt{\frac{2A}{\pi}} e^{-\frac{A}{2}S_{23}^2}.
\end{eqnarray}
Using this, we may define an average spacing, $\overline{S_{23}}$ through the first moment and obtain finally a normalized
spacing distribution in terms of the variable $z = S_{23}/\overline{S_{23}}$:
\begin{equation}\label{eq:9}
p_{cc}(z) = \frac{2}{\pi} e^{-\frac{z^2}{\pi }}.
\end{equation}

Similarly, the spacing distribution, $P_{rc}(S_{12})$ is obtained:
\begin{equation}\label{eq:10}
P_{rc}(S_{12}) = \frac{4A}{\sqrt{3}}S_{12}e^{-\frac{4}{3}S_{12}^2}I_0\left(\frac{2}{3}AS_{12}^2\right).
\end{equation}
Mean spacing turns out to be $\overline{S_{12}} = \frac{3}{8}\sqrt{\frac{\pi}{A}}c$
where $c = ~_{2}F_{1}\left[\frac{3}{4}, \frac{5}{4}, 1, \frac{1}{4}\right] = 1.31112...$ Defining $z = S_{12}/\overline{S_{12}}$,
\begin{eqnarray}\label{eq:11}
p_{rc}(z)&=&\frac{3\sqrt{3}\pi }{16} c^2z\exp \left(-\frac{3\pi }{16} c^2z^2\right)\nonumber \\
&~& I_0\left(\frac{3\pi }{32} c^2z^2\right).
\end{eqnarray}

\begin{figure}[h]
\begin{center}
\includegraphics[scale=1]{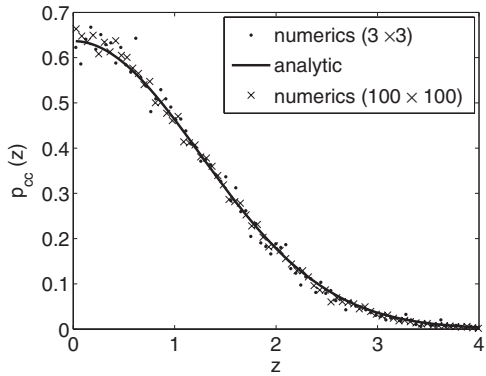}
\end{center}
\caption{Probability distribution of the absolute spacing between the complex conjugate pair of eigenvalues of a
Gaussian ensemble of $3 \times 3$ cyclic matrices. The numerical result obtained by considering 10000 realizations agrees with
the analytic result (\ref{eq:9}). The Gaussian spacing distribution may be interpreted to give an accumulation of eigenvalues
resulting in a maximum at zero spacing, but no tendency to cluster as the first derivative is zero. This is different from a
Poisson distribution. \label{fig:pcc3by3}}
\end{figure}

\begin{figure}[h]
\begin{center}
\includegraphics[scale=1]{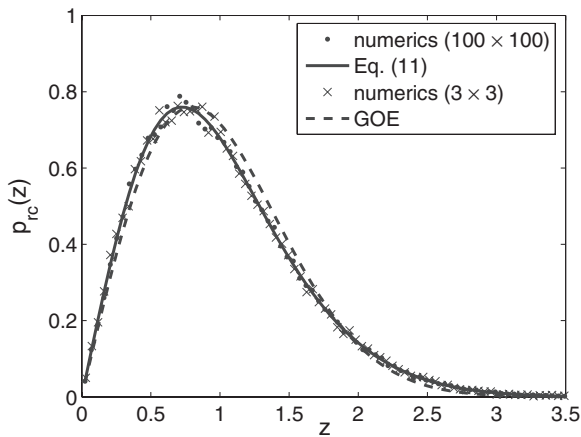}
\end{center}
\caption{Probability distribution of the absolute spacing between a real and a complex eigenvalue of a
Gaussian ensemble of $3 \times 3$ cyclic matrices. The numerical result obtained by considering 10000 realizations for
$3 \times 3$ matrices and 1000 realizations of $100 \times 100$ matrices agrees with
the analytic result (\ref{eq:11}). We observe a linear level repulsion near zero spacing, however the result is distinctly
different from the Wigner surmise for GOE. \label{fig:prc3by3}}
\end{figure}

We can now make following observations :  (i) the Gaussianity of $p_{cc}(z)$ implies that there is no level repulsion among the
complex conjugate pairs, at the same time there is
no attraction, there is no tendency of clustering as in Poissonian spacing distribution; (ii) real and complex eigenvalues display
linear level repulsion. These results are also borne out by the numerical simulations
in Fig. \ref{fig:pcc3by3} and Fig. \ref{fig:prc3by3}.

\begin{figure}[h]
\begin{center}
\includegraphics[scale=1]{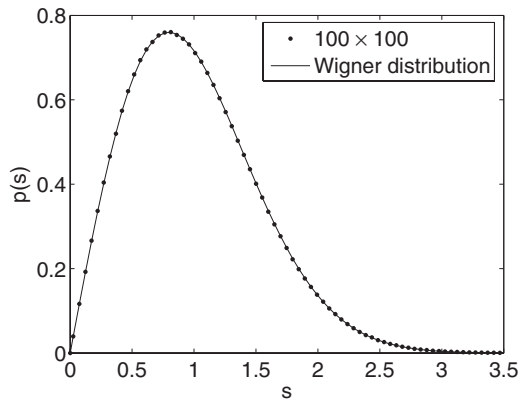}
\end{center}
\caption{ We observe a linear level repulsion between two eigenvalues which are neither real nor complex conjugate pairs for an ensemble of
$100 \times 100$ matrices with
5000 realizations. The agreement with GOE is deceptive; in fact, this suggests that the eigenvalues describe a Poisson process on a plane. \label{fig:prc100by100}}
\end{figure}

For the general case of $N \times N$ matrices, we need to invert (\ref{eq:4}). This inversion leads us to the following relation:
\begin{eqnarray}\label{eq:12}
a_i = \sum_{l} \mbox{\helv S}_{il} E_l
\end{eqnarray}
where {\helv S}$_{il}$ = $\omega ^{(i-1)(N-(l-1))}$ and $\omega = e^{2\pi i/N}$ is a root of unity.  {\helv S} is a symmetric matrix and
{\helv S}$^2 = N\beeta $. Employing these relations, we can find $\sum_{i} a_i^2$, and hence the following result for the JPDF for even $N$:
\begin{eqnarray}\label{eq:13}
P(\{E_i\}) &=& \left( \frac{A}{\pi } \right)^{\frac{N}{2}} \exp \bigg[-A \bigg(E_1^2 + E_{\frac{N}{2}+1}^2 \nonumber \\
&~&~~~~~+ \sum_{i \neq 1, \frac{N}{2}+1}^{N} E_iE_{N+2-i} \bigg)\bigg]
\end{eqnarray}
where $E_1$ and $E_{\frac{N}{2}+1}$ real and the rest of the eigenvalues may be complex. For odd $N$, the above result will hold except that
there will be only one real eigenvalue, $E_1$ and the summation in the second term will extend over all $i$ except 1. Employing this
general result on JPDF, we can now calculate the spacing distributions for the general case. There are three cases : (i) spacing among the
complex conjugate pair of eigenvalues is found to be distributed again as a Gaussian; (ii) spacing between a real and a complex eigenvalue is
distributed according to (\ref{eq:11}); (iii) two complex eigenvalues, $E_j = x_j + iy_j$ and $E_k = x_k + iy_k$ are spaced according to
\begin{equation}\label{eq:14}
p(s) = \frac{\int \prod_{i} d\Re{E_i}d\Im{E_i} P(\{E_i\})\delta (|E_j - E_k| - s)}{\int \prod_{i} d\Re{E_i}d\Im{E_i} P(\{E_i\})},
\end{equation}
which reduces to the following integral on change of variables, $\xi (\eta ) _{\pm} = x(y)_k \pm x(y)_j$
\begin{eqnarray}\label{eq:15}
p(s) &=& \frac{A}{\pi} \int d\xi _- d\eta _- e^{-A(\xi _-^2 + \eta _-^2)}\delta (\sqrt{\xi _-^2 + \eta _-^2} - s) \nonumber \\
&=& \frac{\pi s}{2}\exp \left(-\frac{\pi s^2}{4}\right)
\end{eqnarray}
which is exactly the Wigner distribution (Fig. \ref{fig:prc100by100}).  Let us recall that Wigner's
result holds exactly for $2 \times 2$ real symmetric matrices, it serves as an excellent approximation for $N \times N$ matrices though. We also know
that the spacing distribution for a Poissonian random process in a plane is exactly the same as Wigner surmise. Thus our result proves that the
complex eigenvalues of random cyclic matrices describe such a process. This is a very beautiful, non-intuitive result which brings out yet another characteristic of
RCM.

The eigenfunctions of {\helv M} corresponding to the real eigenvalues ($E_1$ and $E_{\frac{N}{2}+1}$) are also simultaneously eigenfunctions
of ``generalized parity" $\beeta $. However, the eigenfunctions of {\helv M} corresponding to the complex conjugate pair of eigenvalues
are not simultaneously eigenfunctions of $\beeta$. Thus, when these complex eigenvalues occur, ``generalized parity" is said to be
spontaneously broken. Also, the eigenfunctions corresponding to the complex conjugate pair of eigenvalues have zero ${\cal PT} $ - norm.
This is expected from the recent works \cite{bender,ahmed} on ${\cal PT}$-symmetric quantum mechanics.  This observation then fully embeds our findings
into the new random matrix theory developed recently for pseudo-Hermitian
Hamiltonians.  However, we also note that the eigenvectors $\psi _1$ ($\psi _2$) corresponding to complex conjugate eigenvalues, $\lambda $ ($\lambda ^*$)
satisfy orthogonality defined with respect to $\beeta$. Since these results are found for $N \times N$ matrices, we believe that this work extends the random matrix theory in a significant way.
The findings on the spacing distributions have led
us to a linear level repulsion among distinct complex eigenvalues, whereas the spacing between complex-conjugate pair is Gaussian-distributed.

Ginibre orthogonal ensemble with Gaussian distributed real elements has been completely solved only recently \cite{Kanzieper,Peter}. The ensemble of asymmetric random cyclic matrices is a simple nontrivial instance for which all the interesting quantities are analytically obtained in an explicit manner. Such examples play an important role in developing a deeper insight, even when  formal results exist.

Also, we would
like to point out the role played by level repulsion when a system with spectral properties described by RMT approaches equilibrium. In its approach to equilibrium,
the central quantity of interest is the two-time correlation function, the long-time behaviour is decided by the degree of level repulsion as the levels get
closer. We can immediately see \cite{srjpg} that linear level repulsion is related to the exponent, 2 in $t^{-2}$-tail at long times.

It is a great pleasure to thank Bob Dorfman, University of Maryland, College Park, U.S.A. for bringing to our notice the role
played by cyclic matrices in certain models in statistical mechanics, the work of Ted Berlin and Mark Kac \cite{spherical} in particular.


\begin{thebibliography}{99}
\bibitem{note} Note that we are not using the term ``nearest neighbour spacing distribution" because the spacings are in general complex.
\end{thebibliography}
\end{document}